\documentclass[11pt]{article}
\usepackage[letterpaper,margin=1.00in]{geometry}
\usepackage[T1]{fontenc}
\usepackage{amsmath,amssymb,mathtools,bm}
\DeclareMathAlphabet{\mathbf}{OT1}{cmr}{bx}{n}
\usepackage{graphicx,microtype}
\usepackage[square,numbers,sort&compress]{natbib}
\usepackage{xcolor}
\definecolor{linkblue}{RGB}{0,50,232}
\definecolor{equationpink}{RGB}{255,0,127}
\usepackage[colorlinks=true,linkcolor=linkblue,citecolor=linkblue,urlcolor=linkblue]{hyperref}
\makeatletter
\renewcommand{\eqref}[1]{\textup{\hyperref[#1]{\textcolor{equationpink}{\tagform@{\ref*{#1}}}}}}
\makeatother
\numberwithin{equation}{section}
\newcommand{\dd}{\mathrm d}
\newcommand{\GB}{\mathcal L_{\rm GB}}
\newcommand{\E}{\widehat{\mathcal E}}
\newcommand{\doilink}[1]{\href{https://doi.org/#1}{\textcolor{linkblue}{doi:#1}}}

\title{\bfseries Exact Hairy Black Holes in Higher-Curvature\\
Scalar--Tensor Gravity}
\author{Tianhao Wu\thanks{Corresponding author: twu49@illinois.edu}\\
\normalsize Department of Physics, University of Illinois Urbana-Champaign,\\[-2pt]
\normalsize Urbana, Illinois 61801, USA}
\date{}

\begin{document}
\maketitle

\begin{abstract}
We construct exact hairy black holes in Lovelock--Horndeski gravity and establish a relation between horizon topology, scalar hair and covariant charges. Rooted in the dimensional reduction of higher-curvature gravity, this framework connects scalar--tensor dynamics to the low-energy description of heterotic strings. In five dimensions, a single fixed theory admits a continuous planar family and two isolated hyperbolic black holes. The hyperbolic solutions share a locally AdS metric with distinct scalar dressings, while the planar family is Weyl curved and has a freely varying horizon scale. Both branches carry scalar hair regular on the future horizon. The hyperbolic construction extends above four dimensions; compatibility of the displayed planar profile selects five and seven dimensions.
Time translations of the scalar act through a global shift, which enters the complete covariant charge balance. Along the planar family, a changing bulk shift-charge density supplies this balance despite zero radial flux and vanishing horizon charge variation. The hyperbolic roots admit no horizon-scale variation at fixed couplings.
\end{abstract}

\clearpage

\section{Introduction}

Scalar fields and higher-curvature interactions arise together in the low-energy description of string theory. Compactification turns part of the higher-dimensional geometry into dynamical scalars and relates their interactions to those of the metric. Lovelock--Horndeski gravity gives this connection a systematic form. Lovelock's curvature construction and Horndeski's four-dimensional scalar--tensor theory identify complementary ways to retain second-order field equations
\cite{Lovelock1971,Horndeski1974}. Covariant Galileons expose the corresponding derivative cancellations
\cite{NicolisRattazziTrincherini2009,DeffayetEspositoFareseVikman2009},
and the reduction of Lovelock densities gives those scalar interactions a geometric origin \cite{VanAcoleyenVanDoorsselaere2011}.

The heterotic construction of Wu and Stone \cite{WuStone2026} derives a Lovelock--Horndeski branch from the metric--dilaton sector and determines its holographic conformal anomaly and $a$-theorem. The subsequent holographic renormalization analysis supplies the boundary variational problem, finite responses
and source-dependent Ward identities \cite{WuHolographicRenormalization2026}. The black-hole solutions probe these string-derived interactions through regular horizons supported by nontrivial scalar hair.

The integration constants of an exact black-hole solution describe the configurations available to a fixed theory, while its horizon and charges expose how the gravitational symmetries are realized. In Lovelock gravity, the curvature couplings organize distinct black-hole branches \cite{BoulwareDeser1985,Wheeler1986,CrisostomoTroncosoZanelli2000,Cai2002}. With a dynamical scalar, horizon regularity and symmetry inheritance become equally important.
The shift-symmetric no-hair theorem \cite{HuiNicolis2013} and the construction of regular time-dependent scalar hair
\cite{BabichevCharmousis2014,KobayashiTanahashi2014} establish how stationary geometries support scalars linear in time. Scalar
coupling to the Gauss--Bonnet invariant supplies a further connection between hair and curvature \cite{SotiriouZhou2014}. Analytic black holes in reduced Lovelock theories establish the geometric setting for studying these effects together
\cite{CharmousisGouterauxKiritsis2012}.

Here we determine how horizon topology and scalar hair organize the solutions and covariant charges of the Lovelock--Horndeski theory. In five dimensions, one fixed action supports a continuous planar black-brane family and two isolated hyperbolic black holes. The planar geometry is Weyl curved, its scalar gradient is timelike, and its horizon radius remains free. The hyperbolic configurations have the same locally AdS metric but different scalar profiles with constant spacelike gradients. Fixing the action and the scalar normalization makes this distinction physical: a continuous black-hole family and discrete scalar dressings coexist without changing the theory. The common coupling locus follows from an exact algebraic intersection, with two real hyperbolic roots selected by the planar conditions.

The dependence on dimension exposes a further consequence of the coupled equations. Keeping the transverse dimension arbitrary throughout variation and elimination gives a hyperbolic construction for every $D\geq5$, whereas the displayed planar profile is compatible only with $D=5$ and $D=7$. Explicit elimination of the dimension-dependent coefficient equations identifies the origin of this selection.

The scalar symmetry also changes the interpretation of the gravitational charges. A time translation shifts the scalar by a
constant, so the full symmetry of the configuration combines the spacetime flow with a global scalar shift. The complete covariant potentials yield $\dd\mathbf k_\xi=q\,\delta\mathbf j$, including variations of the time slope $q$. On the planar family, a changing bulk shift-charge density supplies the finite-section balance with vanishing radial current flux. The horizon charge variation is zero, while the hyperbolic roots have no horizon-scale tangent at fixed action.
These results connect the geometry of the solution space to the realization of time translations in the covariant variational
problem \cite{Wald1993,IyerWald1994}.

The same backgrounds give concrete bulk configurations for AdS/CFT \cite{Maldacena1998,GubserKlebanovPolyakov1998,Witten1998}. The planar source--temperature relation supplies boundary data for thermal response and transport \cite{SonStarinets2002,IqbalLiu2009,FigueroaPallikaris2020}; the two scalar dressings of one hyperbolic metric separate scalar boundary data from the geometry underlying topological AdS thermodynamics and spherical entanglement \cite{Birmingham1999,ArosTroncosoZanelli2001,Emparan1999,CasiniHuertaMyers2011}.
Reference~\cite{WuCompanionBranes2026} studies dilaton-coupled Lovelock--Horndeski gravity, where the overall exponential factor relates black-brane existence to stationary horizon entropy.

\section{Lovelock--Horndeski gravity}
\label{sec:PHtheory}

We study the shift-symmetric five-dimensional action
\begin{equation}
 S=\frac{1}{16\pi G_5}\int\dd^5x\sqrt{-g}\Big(
 R+\alpha_0X+\alpha_1\GB
 +\alpha_2G^{\mu\nu}\phi_\mu\phi_\nu
 +\alpha_3X\Box\phi+\alpha_4X^2\Big),
 \label{eq:action}
\end{equation}
where
\begin{equation}
 \qquad X=g^{\mu\nu}\phi_\mu\phi_\nu , \qquad  \GB=R_{\mu\nu\rho\sigma}R^{\mu\nu\rho\sigma}
 -4R_{\mu\nu}R^{\mu\nu}+R^2 .
 \label{eq:GBdefinition}
\end{equation}
We take $\phi$ to be dimensionless and $[r]=[L]$.  Then $[q]=L^{-1}$,
$\alpha_0$ is dimensionless, and $\alpha_1,\ldots,\alpha_4$ have dimension $L^2$.
The interactions follow the unweighted heterotic construction of~\cite{WuStone2026,WuHolographicRenormalization2026}; all five couplings are nonzero on the five-dimensional branches below.
For the continuation in section~\ref{sec:PHdimensions} we use this operator density in $D$ dimensions, with the dimension-dependent coupling loci computed explicitly. The common-action construction sets $D=5$.
The potential-free action has an exact shift symmetry, and its derivative interactions support the AdS curvature scale. We compare solutions at fixed couplings and scalar normalization, so their continuous parameters are moduli of one theory.

\subsection{String origin and dimensional coefficient maps}
\label{sec:PHstringorigin}

For a ten-dimensional metric--dilaton parent, write the flat-torus reduction data as
\begin{equation}
\dd\widehat s_{10}^2=e^{2\alpha_{\rm K}\phi}\dd s_D^2
+e^{2\beta_{\rm K}\phi}\dd\bm y_n^2,\qquad
\widehat\Phi=s\phi,\qquad n=10-D.
\label{eq:PHKKansatz}
\end{equation}
The two- and four-derivative sectors carry the respective scalar weights $\exp\{((D-2)\alpha_{\rm K}+n\beta_{\rm K}-2s )\phi\}$
and $\exp\{((D-4)\alpha_{\rm K}+n\beta_{\rm K}-2s)\phi\}$. Their common unweighted representative therefore has
\begin{equation}
\alpha_{\rm K}=0,\qquad n\beta_{\rm K}=2s,\qquad
\alpha_0^{\rm KK}=4s^2-n(n+1)\beta_{\rm K}^2
=-\frac{4s^2}{n}.
\label{eq:PHKKkinetic}
\end{equation}
Here the Laplacian in
$\widehat R=R-2n\beta_{\rm K}\Box\phi-n(n+1)\beta_{\rm K}^2X$ is a total derivative in the unweighted action. For $s=1$, the
$10\to5$ normalization is $\alpha_0=-4/5$, as used at the common black-hole point below. The $10\to7$ normalization is instead
$\alpha_0=-4/3$.

The higher-derivative coefficients are likewise images of a specified parent coefficient frame and reduction:
\begin{equation}
\boldsymbol\alpha^{(D)}
=\mathcal M_{10\to D}(\Theta_{\rm str};n=10-D).
\label{eq:PHKKmap}
\end{equation}
The internal dimension enters the curvature multiplicities and the integration-by-parts coefficients, so this map changes with $D$. Our dimensional black-hole equations determine the supporting lower-dimensional coupling loci. Their intersection with the image of \eqref{eq:PHKKmap} selects the black holes realized by each compactification. This formulation connects the heterotic construction \cite{WuStone2026,WuHolographicRenormalization2026} to the dimension-dependent solution space.

\subsection{Covariant field equations}

We use $X=\nabla_\rho\phi\nabla^\rho\phi$ and normalize the
Euler tensors through the bulk variation
\begin{equation}
 \delta S=\frac{1}{16\pi G_D}\int\dd^Dx\sqrt{-g}
 \left(\mathcal E_{\mu\nu}\delta g^{\mu\nu}
 +\mathcal E_\phi\delta\phi\right).
 \label{eq:covariantvariation}
\end{equation}
The complete metric equation is
\begin{equation}
\begin{aligned}
\mathcal E_{\mu\nu}={}&G_{\mu\nu}
+\alpha_0\left(\nabla_\mu\phi\nabla_\nu\phi
-\frac12g_{\mu\nu}X\right)
+\alpha_1\mathcal H^{\rm GB}_{\mu\nu}\\
&+\alpha_2\Bigl\{
-\frac12R\nabla_\mu\phi\nabla_\nu\phi
+2R_{\rho(\mu}\nabla_{\nu)}\phi\nabla^\rho\phi
+R_{\mu\rho\nu\sigma}\nabla^\rho\phi\nabla^\sigma\phi
-\frac12XG_{\mu\nu}\\
&\qquad
+(\nabla_\mu\nabla_\rho\phi)(\nabla_\nu\nabla^\rho\phi)
-(\Box\phi)\nabla_\mu\nabla_\nu\phi\\
&\qquad
+g_{\mu\nu}\Bigl(
-\frac12(\nabla_\rho\nabla_\sigma\phi)
          (\nabla^\rho\nabla^\sigma\phi)
+\frac12(\Box\phi)^2
-R_{\rho\sigma}\nabla^\rho\phi\nabla^\sigma\phi
\Bigr) \Bigr\}\\
&+\alpha_3\Bigl(
(\Box\phi)\nabla_\mu\phi\nabla_\nu\phi
-2\nabla_{(\mu}\phi\,
  (\nabla_{\nu)}\nabla_\rho\phi)\nabla^\rho\phi
+g_{\mu\nu}\nabla^\rho\phi\nabla^\sigma\phi\,
  \nabla_\rho\nabla_\sigma\phi\Bigr)\\
&+\alpha_4\left(
2X\nabla_\mu\phi\nabla_\nu\phi
-\frac12g_{\mu\nu}X^2\right)=0,
\end{aligned}
\label{eq:metricEOM}
\end{equation}
where the Gauss--Bonnet tensor is
\begin{equation}
\begin{aligned}
\mathcal H^{\rm GB}_{\mu\nu}={}&
2\Bigl(RR_{\mu\nu}-2R_{\mu\rho}R_\nu{}^\rho
-2R^{\rho\sigma}R_{\mu\rho\nu\sigma}
+R_\mu{}^{\rho\sigma\lambda}R_{\nu\rho\sigma\lambda}\Bigr)
-\frac12g_{\mu\nu}\GB.
\end{aligned}
\label{eq:covarianttensors}
\end{equation}
The complete scalar equation is
\begin{equation}
\begin{aligned}
\mathcal E_\phi={}&
-2\alpha_0\Box\phi
-2\alpha_2G^{\mu\nu}\nabla_\mu\nabla_\nu\phi\\
&+2\alpha_3\Bigl(
(\nabla_\mu\nabla_\nu\phi)(\nabla^\mu\nabla^\nu\phi)
-(\Box\phi)^2
+R_{\mu\nu}\nabla^\mu\phi\nabla^\nu\phi\Bigr)\\
&-4\alpha_4\Bigl(
2\nabla^\mu\phi\nabla^\nu\phi\,
  \nabla_\mu\nabla_\nu\phi
+X\Box\phi\Bigr)=0.
\end{aligned}
\label{eq:scalarEOM}
\end{equation}
Both equations contain derivatives of at most second order.
Diffeomorphism invariance gives the off-shell identity
\begin{equation}
 2\nabla^\mu\mathcal E_{\mu\nu}
 +\mathcal E_\phi\nabla_\nu\phi=0.
 \label{eq:covariantNoether}
\end{equation}
\subsection{Exact four-function reduction}

We take the homogeneous future-ingoing Ansatz
\begin{equation}
 \dd s^2=-n(r)^2f(r)\dd v^2+2n(r)\dd v\dd r
 +\Sigma(r)^2\dd\Sigma_{3,k}^2,
 \qquad \phi=qv+p(r),\qquad k=0,-1,
 \label{eq:seed}
\end{equation}
with $R_{ij}(\gamma_k)=2k\gamma_{ij}$. Its kinetic density is
\begin{equation}
 X=\frac{2qp'}n+f(p')^2 .
 \label{eq:X}
\end{equation}
The complete configuration realizes stationarity through the metric Killing vector and the exact scalar shift:
\begin{equation}
 \mathcal L_{\partial_v}g_{\mu\nu}=0,\qquad
 \mathcal L_{\partial_v}\phi=q.
 \label{eq:stationaryshift}
\end{equation}
Consequently $X$, every operator in the action, and the field equations are independent of $v$. The homogeneous fiber has $\nabla_i\phi=0$ and, after $n=1$ and $\Sigma=r$, the scalar Hessian
\begin{equation}
 \nabla_i\nabla_j\phi=r\bigl(q+fp'\bigr)\gamma_{ij}.
 \label{eq:fiberHessian}
\end{equation}

For an explicit action-level reduction, leave the fiber dimension $\nu=D-2$ unspecified and write $\dd s_D^2=h_{AB}\dd x^A\dd x^B+\Sigma^2\gamma_{ij}\dd x^i\dd x^j$. We normalize the maximally symmetric fiber by $R_{ij}(\gamma)=(\nu-1)k\gamma_{ij}$.  Let $D_A$ be the covariant derivative of $h_{AB}$ and define
\begin{equation}
 S_{AB}=D_AD_B\Sigma,\qquad B_\Sigma=D^AD_A\Sigma,\qquad
 U=D_A\Sigma D^A\Sigma,\qquad
 Q_\Sigma=(\nu-1)k-\Sigma B_\Sigma-(\nu-1)U.
 \label{eq:warpedblocks}
\end{equation}
The complete curvature reduction is
\begin{equation}
\begin{aligned}
R_D={}&R_h+\frac{\nu(\nu-1)(k-U)}{\Sigma^2}
-\frac{2\nu B_\Sigma}{\Sigma},\\
R_{\mu\nu}R^{\mu\nu}={}&
\left(R^{(h)}_{AB}-\frac{\nu}{\Sigma}S_{AB}\right)
\left(R_{(h)}^{AB}-\frac{\nu}{\Sigma}S^{AB}\right)
+\frac{\nu Q_\Sigma^2}{\Sigma^4},\\
R_{\mu\nu\rho\sigma}R^{\mu\nu\rho\sigma}={}&R_h^2
+\frac{4\nu}{\Sigma^2}S_{AB}S^{AB}
+\frac{2\nu(\nu-1)}{\Sigma^4}(k-U)^2.
\end{aligned}
\label{eq:warpedcurvature}
\end{equation}
With $V_A=D_A\phi$ and $X=V_AV^A$, the remaining two derivative contractions are
\begin{equation}
\begin{aligned}
H_\phi\equiv\Box_D\phi&=D^AD_A\phi
+\frac{\nu}{\Sigma}D_A\Sigma V^A,\\
Z_\phi\equiv G^{\mu\nu}\phi_\mu\phi_\nu
&=\left(R^{(h)}_{AB}-\frac{\nu}{\Sigma}S_{AB}\right)V^AV^B
-\frac12R_DX.
\end{aligned}
\label{eq:warpedscalar}
\end{equation}
Equations~\eqref{eq:warpedcurvature} and \eqref{eq:warpedscalar} give the explicit reduced density per unit fiber volume:
\begin{equation}
{\cal L}_{\rm red}=n\Sigma^\nu\Big(
R_D+\alpha_0X+\alpha_1\GB+\alpha_2Z_\phi
+\alpha_3XH_\phi+\alpha_4X^2\Big),
\qquad \GB=R_{\mu\nu\rho\sigma}^2-4R_{\mu\nu}^2+R_D^2.
\label{eq:reducedLagrangian}
\end{equation}
The two-dimensional metric contains the lapse and the radial blackening function, while $\Sigma$ determines the physical size of the homogeneous fiber.  Keeping these functions independent retains the metric equations associated with variations of both the normal geometry and the transverse volume.  The scalar equation follows from varying its radial profile, with the constant time slope specifying the stationary-up-to-shift ansatz.
We vary \eqref{eq:reducedLagrangian} with respect to $n,f,\Sigma,p$ before imposing either $\Sigma=r$ or a lapse condition.
We denote the reduced Euler derivatives by $\E_Q$.
Including second radial derivatives, the four reduced equations are
\begin{equation}
 \E_Q=\frac{\partial{\cal L}_{\rm red}}{\partial Q}
 -\frac{\dd}{\dd r}\frac{\partial{\cal L}_{\rm red}}{\partial Q'}
 +\frac{\dd^2}{\dd r^2}
 \frac{\partial{\cal L}_{\rm red}}{\partial Q''}=0,
 \qquad Q\in\{n,f,\Sigma,p\}.
 \label{eq:Euleroperator}
\end{equation}
Their exact relation to the covariant equations is
\begin{equation}
\begin{aligned}
\E_n&=-\frac{2\Sigma^\nu}{n}\mathcal E_{vr},&
\E_f&=n\Sigma^\nu\mathcal E_{rr},\\
\E_\Sigma&=-2n\Sigma^{\nu-3}\gamma^{ij}\mathcal E_{ij},&
\E_p&=n\Sigma^\nu\mathcal E_\phi .
\end{aligned}
\label{eq:covariantReducedBridge}
\end{equation}
The choices $\Sigma=r$ and $n=1$ select the areal radius and the constant lapse of the exact branches after variation; a constant rescaling of $v$ fixes the lapse normalization. The scalar equation is $\mathcal E_\phi=-\nabla_\mu J^\mu=0$, equivalently $\partial_r(n\Sigma^\nu J^r)=0$, where
\begin{equation}
 J^\mu=2\alpha_0\phi^\mu+2\alpha_2G^{\mu\nu}\phi_\nu
 +2\alpha_3\left((\Box\phi)\phi^\mu-\phi^{\mu\nu}\phi_\nu\right)
 +4\alpha_4X\phi^\mu .
 \label{eq:current}
\end{equation}
Here $\phi^{\mu\nu}=\nabla^\mu\nabla^\nu\phi$.  Exact simplification of \eqref{eq:Euleroperator} cancels all third and fourth derivatives through the Lovelock--Horndeski structure.  The remaining $vv$ equation is then constrained by \eqref{eq:covariantNoether}; it is evaluated independently on the explicit P backgrounds below.

For a branch $B$, we expand the four symmetry-reduced residuals $\E_A|_B$, $A=n,f,\Sigma,p$, in independent radial monomials and denote the primitive coefficients by $\widehat{\mathcal C}^{B,A}_{\nu}$.  Their independence gives
\begin{equation}
 \E_A|_B=0\quad\Longleftrightarrow\quad
 \widehat{\mathcal C}^{B,A}_{\nu}=0,
 \qquad
 I_B=\left\langle\widehat{\mathcal C}^{B,A}_{\nu}\right\rangle .
 \label{eq:ideal}
\end{equation}
For each radial profile considered below, we determine its coupling locus in the chamber with nonzero couplings, horizon scale and scalar scale.
Saturating by their product implements this chamber algebraically.
Zero normal forms establish the equivalence between the finite ideal and the four radial Euler expressions, and solving the ideal fixes the coupling locus of each profile.

\section{Planar black brane Solution}

\subsection{Horizon regularity and coefficient ideal}
For $k=0$ and $n=1$, a finite radial slope with $q=0$ gives the leading simple-horizon lapse equation $\E_n^{(0)}=-3r_h^2F'(r_h)=0$. For $r_h>0$, this is the extremal condition.
We therefore take $q\ne0$ and impose horizon regularity through the planar profiles
\begin{equation}
 F=\frac{r^2}{L^2}\left(1-\left(\frac{r_h}{r}\right)^m\right),
 \qquad p'=c\,\frac{1-(r_h/r)^z}{F} .
 \label{eq:Ptrial}
\end{equation}
For the scalar configuration $\phi=qv+p(r)$, the radial component of its gradient is $\nabla^r\phi=q+Fp'$. We introduce
\begin{equation}
 Y(r)=1+\frac{F(r)p'(r)}q,\qquad
 X=\frac{q^2(Y^2-1)}F .
 \label{eq:Y}
\end{equation}
Finiteness of $p'$ at a simple future horizon requires $Y(r_h)=1$.
We impose $Y\to0$ at infinity, fixing the leading radial falloff of the scalar. The resulting boundary conditions are
\begin{equation}
 Y(r_h)=1,\qquad Y(\infty)=0,\qquad
 p'(r_h)=\frac{qY'(r_h)}{F'(r_h)},\qquad
 p'=-\frac{qL^2}{r^2}+o(r^{-2}).
 \label{eq:Pboundary}
\end{equation}
The near-horizon expansion of \eqref{eq:Ptrial}, with
$\rho=r-r_h$, gives
\begin{equation}
 F=\frac{mr_h}{L^2}\rho+O(\rho^2),\qquad
 p'=\frac{czL^2}{mr_h^2}+O(\rho),\qquad
 X_h=\frac{2qczL^2}{mr_h^2}.
 \label{eq:Phorizontrial}
\end{equation}
Thus both the scalar gradient in ingoing coordinates and its kinetic invariant remain finite at the horizon. Since
$Y(\infty)=1+c/q$, the asymptotic condition fixes $c=-q$ and yields $Y=(r_h/r)^z$.

We now determine the allowed exponents and couplings from the five-dimensional field equations. Within the integer-power family
$1\leq m,z\leq5$, compatibility of all four radial equations with nonzero $\alpha_0,\ldots,\alpha_4$ uniquely selects
$(m,z)=(2,1)$ and fixes the dimensionless coupling relations.
Appendix~\ref{app:Pcoefficients} derives these relations and establishes the incompatibility of the remaining exponent pairs
by exact algebraic elimination. The resulting branch is
\begin{equation}
 F=\frac{r^2}{L^2}\left(1-\frac{r_h^2}{r^2}\right),\qquad
 p'=-\frac{qL^2}{r(r+r_h)} .
 \label{eq:Pseed}
\end{equation}
The future-regular geometry and scalar are therefore
\begin{equation}
 \dd s^2=-F(r)\dd v^2+2\dd v\dd r+r^2\dd\bm{x}_3^2,\qquad
 \phi_P=\phi_0+qv+\frac{qL^2}{r_h}\log\frac{r+r_h}{r} .
 \label{eq:P}
\end{equation}

Introduce
\begin{equation}
 (b_0,b_1,b_2,b_3,b_4)=\left(
 \frac{\alpha_0L^4q^2}{r_h^2},\frac{\alpha_1}{L^2},
 \frac{\alpha_2L^2q^2}{r_h^2},\frac{\alpha_3L^4q^3}{r_h^3},
 \frac{\alpha_4L^6q^4}{r_h^4}\right).
 \label{eq:Pdimensionless}
\end{equation}
Specializing the dimension-independent reduction to a three-dimensional horizon gives twelve nonzero coefficient equations after dropping the repeated lapse row. To extract these coefficients from the four equations, take fiber dimension $\nu$ and multiply the uncollected residuals by the nonzero factors
\begin{equation}
\begin{aligned}
\mathcal P_n&=-2L^4(r+r_h)r^{4-\nu}\E_n,&
\mathcal P_f&=-\frac{2(r+r_h)^2r^{4-\nu}}{L^2q^2}\E_f,\\
\mathcal P_\Sigma&=\frac{2L^4r^{5-\nu}}{\nu}\E_\Sigma,&
\mathcal P_p&=\frac{L^2r^{4-\nu}}q\E_p .
\end{aligned}
\label{eq:Pscaledresiduals}
\end{equation}
At $\nu=3$, coefficient extraction in $r$ and the variables \eqref{eq:Pdimensionless}, followed by removal of repetitions and nonzero overall numerical factors, gives the following nine distinct polynomials.
They generate the same zero-dimensional ideal as all twelve rows:
\begin{equation}
\begin{split}
\bm c_P=\{&3b_2-2b_3+b_4,
3b_2-4b_3+3b_4,
-6+b_0+24b_1+6b_2-2b_3,\\
&6+b_0-24b_1+6b_2,
2b_1-1,
3b_2-b_3,
b_0+3b_2,\\
&-2+b_0+8b_1+2b_2,
b_0+6b_2-b_3\}.
\end{split}
\label{eq:Prows}
\end{equation}
Exact lexicographic elimination gives
\begin{equation}
 G_P=\left\{b_0+6,\ b_1-\frac12,\ b_2-2,\ b_3-6,\ b_4-6\right\}.
 \label{eq:Pideal}
\end{equation}
The basis $G_P$ follows directly by linear elimination in \eqref{eq:Prows}.
The fifth, sixth, and seventh rows fix $b_1=1/2$, $b_3=3b_2$, and $b_0=-3b_2$.  The eighth row then gives $b_2=2$, and the first fixes $b_4=6$; the other rows vanish at this point.  This shows explicitly why the overdetermined radial system selects a single dimensionless coupling point. Restoring the dimensional scales gives the coupling orbit
\begin{align}
 \alpha_0&=-\frac{6r_h^2}{L^4q^2},&
 \alpha_1&=\frac{L^2}{2},&
 \alpha_2&=\frac{2r_h^2}{L^2q^2},\notag\\
 \alpha_3&=\frac{6r_h^3}{L^4q^3},&
 \alpha_4&=\frac{6r_h^4}{L^6q^4}.
 \label{eq:Porbit}
\end{align}
Eliminating $L,q,r_h$ gives the fixed-action locus
\begin{equation}
 \alpha_0=-\frac{3\alpha_2}{2\alpha_1},\qquad
 \alpha_4=\frac{3\alpha_2^2}{4\alpha_1},\qquad
 4\alpha_1\alpha_3^2=9\alpha_2^3 .
 \label{eq:Pintrinsic}
\end{equation}
Direct substitution of \eqref{eq:Pseed} and \eqref{eq:Porbit} into the four pre-gauge equations \eqref{eq:Euleroperator} gives
\begin{equation}
 (\E_n,\E_f,\E_\Sigma,\E_p)\big|_P=(0,0,0,0).
 \label{eq:Preducedclosure}
\end{equation}
An independent component substitution into the covariant equations
\eqref{eq:metricEOM} and \eqref{eq:scalarEOM} gives
\begin{equation}
 \left.\mathcal E_{\mu\nu}\right|_P=0,\qquad
 \left.\mathcal E_\phi\right|_P=0,
 \label{eq:Pcovariantclosure}
\end{equation}
including the $vv$ component.
At a fixed point of this locus, $L^2=2\alpha_1$ and $q=3\alpha_2r_h/(2\alpha_1\alpha_3)$. The horizon radius is therefore
the continuous modulus of one action, with the boundary scalar slope tracking the temperature.

\subsection{Geometry, current, and boundary ray}

The horizon data are
\begin{equation}
 F'(r_h)=\frac{2r_h}{L^2},\qquad
 T_P=\frac{r_h}{2\pi L^2},\qquad
 p'(r_h)=-\frac{qL^2}{2r_h^2} .
 \label{eq:PT}
\end{equation}
The curvature and scalar invariants are finite for $r\ge r_h$:
\begin{align}
 R&=-\frac{2(10r^2-3r_h^2)}{L^2r^2},&
 C_{\mu\nu\rho\sigma}C^{\mu\nu\rho\sigma}
 &=\frac{2r_h^4}{L^4r^4},\notag\\
 R_{\mu\nu}R^{\mu\nu}
 &=\frac{4(20r^4-12r^2r_h^2+3r_h^4)}{L^4r^4},&
 R_{\mu\nu\rho\sigma}R^{\mu\nu\rho\sigma}
 &=\frac{4(10r^4-6r^2r_h^2+3r_h^4)}{L^4r^4},\notag\\
 X&=-\frac{L^2q^2}{r^2},&
 \Box\phi&=\frac{2qr_h}{r^2},\notag\\
 \GB&=\frac{24(5r^2-3r_h^2)}{L^4r^2},&
 G^{\mu\nu}\phi_\mu\phi_\nu
 &=-\frac{3q^2(2r^2-r_h^2)}{r^4},\notag\\
 X\Box\phi&=-\frac{2L^2q^3r_h}{r^4},&
 X^2&=\frac{L^4q^4}{r^4}.
 \label{eq:Pinvariants}
\end{align}
The metric is Weyl curved and approaches AdS$_5$. The exact shift current is
\begin{equation}
 J_P^\mu=\left(-\frac{12r_h^2}{L^2qr^2},0,0,0,0\right),\qquad
 \nabla_\mu J_P^\mu=0,\qquad J_P^r=0,
 \label{eq:Pcurrent}
\end{equation}
with norm
\begin{equation}
 J_{P\mu}J_P^\mu=-\frac{144(r^2-r_h^2)r_h^4}{L^6q^2r^4}.
 \label{eq:Pcurrentnorm}
\end{equation}
It is finite throughout the exterior and becomes null on the horizon. At fixed action, the source--temperature ray is
\begin{equation}
 \frac{q}{T_P}=\frac{6\pi\alpha_2}{\alpha_3}.
 \label{eq:PsourceT}
\end{equation}
The leading boundary scalar datum is linear in time. Its radial expansion in static coordinates follows directly from the ingoing solution. In the exterior $r>r_h$, introduce the static time by
\begin{equation}
v=t+r_*(r),\qquad
r_*(r)=\frac{L^2}{2r_h}\log\frac{r-r_h}{r+r_h},\qquad
r_*(\infty)=0 .
\label{eq:Pstatictime}
\end{equation}
Substitution into \eqref{eq:P} gives
\begin{equation}
\begin{aligned}
\phi_P
&=\phi_0+qt+\frac{qL^2}{2r_h}
\log\left(1-\frac{r_h^2}{r^2}\right)\\
&=\phi_0+qt-\frac{qL^2r_h}{2r^2}+O(r^{-4})
\qquad (r\longrightarrow\infty).
\end{aligned}
\label{eq:Pstatictail}
\end{equation}
Thus the $r^{-1}$ term in the ingoing profile is induced by the change of time coordinate; at fixed $t$, the radial tail starts at $r^{-2}$.
The source--temperature relation \eqref{eq:PsourceT}, the metric falloff, and $J_P^r=0$ supply the asymptotic data for holographic response. We use fixed EF coordinates $(v,r)$ for the charge variations in section~\ref{sec:charges}; the corresponding change to static coordinates includes the $r_h$ dependence of $r_*(r)$.

\section{Hyperbolic constant-kinetic black hole solution}

Take $k=-1$ and $F=r^2/R_H^2-1$. Imposing $X=q^2$ gives the scalar quadratic,
\begin{equation}
 F(p')^2+2qp'-q^2=0,\qquad
 p'=\frac{q(-1\pm\sqrt{1+F})}{F} .
 \label{eq:Hroots}
\end{equation}
The two roots simplify to
\begin{equation}
 p'_+=\frac{qR_H}{r+R_H},\qquad
 p'_-=-\frac{qR_H}{r-R_H},\qquad
 \phi_H=\phi_0+qv+qR_H\log\frac{r+R_H}{R_H},
 \label{eq:Hscalar}
\end{equation}
where $p'_+(R_H)=q/2$ is future regular and $p'_-$ has a pole at the future horizon.  The displayed scalar is the integral of the plus root, and the solution is
\begin{equation}
 \dd s^2=-\left(\frac{r^2}{R_H^2}-1\right)\dd v^2
 +2\dd v\dd r+r^2\dd H_3^2,\qquad X=q^2 .
 \label{eq:H}
\end{equation}

Define $\chi=qR_H$ and $L_H^2=R_H^2$. All four Euler equations reduce to two polynomials,
\begin{align}
 M&=24\alpha_1-6\alpha_2\chi^2+\alpha_4\chi^4
 +L_H^2(\alpha_0\chi^2-12),\notag\\
 {\cal S}&=\frac{\alpha_0L_H^2}{2}+3\alpha_2+2\alpha_3\chi
 +\alpha_4\chi^2 .
 \label{eq:MS}
\end{align}
Their exact residual factorization is
\begin{equation}
 \E_n=\frac{r^3M}{R_H^4}-\frac{4q^2r^3{\cal S}}{R_H(r+R_H)},\quad
 \E_f=\frac{2q^2r^3{\cal S}}{(r+R_H)^2},\quad
 \E_\Sigma=\frac{3r^2M}{R_H^4},\quad
 \E_p=-\frac{16qr^3{\cal S}}{R_H^3} .
 \label{eq:Hresiduals}
\end{equation}
Thus $M={\cal S}=0$ is necessary and sufficient in the regular chamber.
On the H background,
\begin{equation}
 \nabla_\mu\nabla_\nu\phi=\frac{q}{R_H}
 \left(g_{\mu\nu}-\frac{\nabla_\mu\phi\nabla_\nu\phi}{q^2}\right),
 \qquad \nabla^\rho\phi\,\nabla_\mu\nabla_\rho\phi=0 .
 \label{eq:Hhessianidentity}
\end{equation}
Substitution into the covariant equations gives the following tensor expressions, in agreement with \eqref{eq:covariantReducedBridge}:
\begin{equation}
\left.\mathcal E_{\mu\nu}\right|_H
=-\frac{M}{2R_H^4}g_{\mu\nu}
+\frac{2{\cal S}}{R_H^2}\nabla_\mu\phi\nabla_\nu\phi,\qquad
\left.\mathcal E_\phi\right|_H
=-\frac{16q}{R_H^3}{\cal S}.
\label{eq:Hcovariantclosure}
\end{equation}
Solving these equations for $\alpha_0$ and $\alpha_4$ gives
\begin{align}
 \alpha_0&=\frac{-48\alpha_1+2R_H^2
 (12+9\alpha_2q^2+2\alpha_3q^3R_H)}{q^2R_H^4},\notag\\
 \alpha_4&=\frac{24\alpha_1-4R_H^2
 (3+3\alpha_2q^2+\alpha_3q^3R_H)}{q^4R_H^4} .
 \label{eq:Horbit}
\end{align}
For $R_H>0$ and $q\ne0$, one fixed action generically selects isolated values of $(L_H^2,\chi)$.  The relevant Jacobian is
\begin{equation}
 \Delta_H=\det\frac{\partial(M,{\cal S})}{\partial(L_H^2,\chi)}
 =2(\alpha_0\chi^2-12)(\alpha_3+\alpha_4\chi)
 -\alpha_0\chi\left(-6\alpha_2+2\alpha_4\chi^2+\alpha_0L_H^2\right).
 \label{eq:Hjacobian}
\end{equation}

The horizon data and active operators are
\begin{equation}
 r_h=R_H,\qquad T_H=\frac1{2\pi R_H},\qquad
 \left(X,\GB,G^{\mu\nu}\phi_\mu\phi_\nu,X\Box\phi,X^2\right)_H
 =\frac1{L_H^4}\left(\chi^2L_H^2,120,6\chi^2,4\chi^3,\chi^4\right).
 \label{eq:Hactivity}
\end{equation}
The total shift current vanishes,
\begin{equation}
 J_H^\mu=0.
 \label{eq:Hcurrent}
\end{equation}
Meanwhile $R_{\mu\nu}=-4g_{\mu\nu}/R_H^2$ and
$C_{\mu\nu\rho\sigma}=0$.  The metric is locally AdS$_5$, while the scalar is nontrivial and all five operator densities in \eqref{eq:Hactivity} are nonzero. A cocompact quotient $H^3/\Gamma$ produces a finite-volume horizon.
The Hessian identity \eqref{eq:Hhessianidentity} organizes the interactions into a vanishing total shift current and the two metric tensor structures in \eqref{eq:Hcovariantclosure}, while each local operator density remains nonzero.

\section{Planar and hyperbolic black holes in one fixed theory}

Let $\ell_*$ be a reference length and set $\bar\alpha_i=\alpha_i/\ell_*^2$ for $i=1,\ldots,4$.  We use the field
normalization for which $\alpha_2=\ell_*^2$.  The common dimensionless point is
\begin{equation}
 (\alpha_0,\bar\alpha_1,\bar\alpha_2,
 \bar\alpha_3,\bar\alpha_4)
 =\left(-\frac45,\frac{15}{8},1,\frac{\sqrt{30}}5,\frac25\right).
 \label{eq:commonaction}
\end{equation}
For P, it is realized by
\begin{equation}
 L^2=\frac{15}{4}\ell_*^2,\qquad
 q_P=\frac{4r_h}{\sqrt{30}\,\ell_*^2},\qquad r_h>0,
 \label{eq:Pcommon}
\end{equation}
so the same action contains a continuous thermal ray. Substitution of
\eqref{eq:commonaction} into the H equations gives
\begin{equation}
 \frac{L_H^2}{\ell_*^2}=\frac{15}{2}+\sqrt{30}\,\chi+\chi^2,
 \label{eq:HcommonL}
\end{equation}
and
\begin{equation}
 2\chi^4+4\sqrt{30}\,\chi^3+120\chi^2
 +60\sqrt{30}\,\chi+225=0.
 \label{eq:Hcommonpoly}
\end{equation}
Indeed, with $y=\chi/\sqrt{30}$ and $u=2y+1$, its left-hand side is
\begin{equation}
 \frac{225}{2}\left(u^4+2u^2-1\right).
 \label{eq:Hcommonfactor}
\end{equation}
Only $u^2=\sqrt2-1$ is nonnegative.  The quartic therefore has exactly two
real roots,
\begin{equation}
 \chi_\pm=\frac{\sqrt{30}}2
 \left[-1\pm\sqrt{\sqrt2-1}\right],\qquad
 \frac{L_H^2}{\ell_*^2}=\frac{15}{2}(\sqrt2-1)>0.
 \label{eq:Hcommonroots}
\end{equation}
At these roots,
\begin{equation}
 \Delta_H(\chi_\pm)=\mp\frac{24\sqrt{15}}5\,\ell_*^2
 \sqrt{\sqrt2-1}\ne0,
 \label{eq:Hcommondelta}
\end{equation}
so both H solutions are isolated.

The two roots have the same metric radius and distinct scalar slopes; they therefore define different complete configurations.

P has $X<0$, nonzero temporal current, a relative $r^{-2}$ metric mode, and a source--temperature ray. H has constant $X>0$, vanishing total current, exact local AdS geometry, and an algebraically selected ratio $q/T_H=2\pi\chi_\pm$. Thus one fixed action supports a continuous planar family and two discrete hyperbolic solutions, distinguished by horizon topology, scalar kinematics and boundary data. The planar time slope varies with temperature, and the hyperbolic roots fix both quantities algebraically.

\section{Dimensional continuation and selection}
\label{sec:PHdimensions}

We now determine the dimensional continuation of both radial mechanisms and their supporting coupling loci in the Lovelock--Horndeski operator basis. Let $D=\nu+2$, with $\nu$ the horizon dimension.

The hyperbolic fields retain their form,
\begin{equation}
 F_H=\frac{r^2}{R^2}-1,\qquad
 \phi_H=\phi_0+qv+qR\log\frac{r+R}{R},\qquad
 X=q^2,\qquad \chi=qR.
 \label{eq:HDfields}
\end{equation}
For arbitrary $\nu$, the four reduced equations factor as
\begin{equation}
\left(\E_n,\E_f,\E_\Sigma,\E_p\right)_H=
\left(
\frac{r^\nu M_D}{R^4}-\frac{4q^2r^\nu{\cal S}_D}{R(r+R)},
\frac{2q^2r^\nu{\cal S}_D}{(r+R)^2},
\frac{\nu r^{\nu-1}M_D}{R^4},
-\frac{4(\nu+1)qr^\nu{\cal S}_D}{R^3}
\right),
\label{eq:HDresiduals}
\end{equation}
where
\begin{equation}
\begin{aligned}
M_D={}&\alpha_1\nu(\nu+1)(\nu-1)(\nu-2)-\nu(\nu+1)R^2
-\frac{\nu(\nu+1)}2\alpha_2\chi^2 +\alpha_0R^2\chi^2+\alpha_4\chi^4,\\
{\cal S}_D={}&\frac{\alpha_0R^2}{2}
+\frac{\nu(\nu+1)}4\alpha_2
+\frac{\nu+1}{2}\alpha_3\chi+\alpha_4\chi^2.
\end{aligned}
\label{eq:HDpolynomials}
\end{equation}
For $R>0$ and $q\ne0$, these equations are equivalent to $M_D={\cal S}_D=0$.  If $M_0$ and ${\cal S}_0$ denote the parts of
$M_D$ and ${\cal S}_D$ independent of $(\alpha_0,\alpha_4)$, the exact coupling orbit is
\begin{equation}
 \alpha_0=\frac{2(-M_0+{\cal S}_0\chi^2)}{R^2\chi^2},\qquad
 \alpha_4=\frac{M_0}{\chi^4}-\frac{2{\cal S}_0}{\chi^2},\qquad
 \det\frac{\partial(M_D,{\cal S}_D)}{\partial(\alpha_0,\alpha_4)}
 =\frac12R^2\chi^4\ne0.
 \label{eq:HDorbit}
\end{equation}
Thus H continues for $D\geq5$ with all five interactions contributing to the field equations. At $D=4$, the unweighted Gauss--Bonnet term is topological, as exhibited by its overall factor of $D-4$ in the metric equation.

The planar mechanism is dimension selective.  Keep
\begin{equation}
 F_P=\frac{r^2-r_h^2}{L^2},\qquad
 Y=\frac{r_h}{r},\qquad
 p'=-\frac{qL^2}{r(r+r_h)}.
 \label{eq:PDfields}
\end{equation}
Exact elimination of the thirteen $\nu$-dependent Laurent coefficients gives
\begin{equation}
 \nu(\nu+1)(\nu-3)(\nu-5)=0.
 \label{eq:Pdimensionselection}
\end{equation}
For physical $\nu\geq2$, only $\nu=3,5$, or $D=5,7$, survive.  In the dimensionless coordinates of \eqref{eq:Pdimensionless}, both are described by
\begin{equation}
(b_0,b_1,b_2,b_3,b_4)=\left(
-\frac{\nu(\nu-1)^2}{2},
\frac1{(\nu-1)(\nu-2)},
\nu-1,\nu(\nu-1),\frac{\nu(\nu-1)(\nu+1)}4
\right).
\label{eq:Pdimensionorbit}
\end{equation}
The coefficient equations in Appendix~\ref{app:Pcoefficients} make the elimination explicit. For $\nu>2$, the Jacobian
determinant of the five selected coefficient equations is
\begin{equation}
\det\frac{\partial(c_{n,4},c_{f,1},c_{f,2},c_{f,0},c_{n,2})}
{\partial(b_0,b_1,b_2,b_3,b_4)}
=128\nu^2(\nu+1)(\nu-1)(\nu-2)\ne0.
\label{eq:Ppivotminor}
\end{equation}
These five independent equations uniquely determine \eqref{eq:Pdimensionorbit}; at $\nu=2$, the remaining lapse coefficient is instead the nonzero constant $12$.
After this substitution, all rows vanish except for repetitions of the three
remainders
\begin{equation}
 \frac12\nu(\nu-3)(\nu-5),\qquad
 \frac12\nu(\nu-3)(\nu-5),\qquad
 -\frac12(\nu-3)(\nu-4)(\nu-5),
 \label{eq:Pdimensionremainders}
\end{equation}
which select $D=5,7$ for the planar profile \eqref{eq:PDfields}.  The second allowed point is
\begin{equation}
 D=7:\quad
 (b_0,b_1,b_2,b_3,b_4)=\left(-40,\frac1{12},4,20,30\right),
 \label{eq:P7orbit}
\end{equation}
and direct substitution gives
$\mathcal E_{\mu\nu}=0=\mathcal E_\phi$ as well as all four reduced residuals. An independent classification over $\nu=2,\ldots,7$ and integer blackening and interpolation powers $1,\ldots,5$ selects exactly $(\nu,m,z)=(3,2,1)$ and $(5,2,1)$.

For a ten-dimensional interpretation of either allowed point, $n=10-D$ must be used in \eqref{eq:PHKKmap}. In the unweighted
normalization \eqref{eq:PHKKkinetic}, compatibility of the planar kinetic coefficient fixes
\begin{equation}
\frac{q^2L^4}{r_h^2}
=\frac{\nu(\nu-1)^2(10-D)}{8s^2},\qquad
D=\nu+2\in\{5,7\}.
\label{eq:PHKKplanarscale}
\end{equation}
Equation~\eqref{eq:PHKKplanarscale} fixes the scalar normalization of the planar orbit for each compactification dimension. The full compactification condition is the intersection of this orbit with the dimension-dependent image of \eqref{eq:PHKKmap}.

The dimensional distinction follows from the curvature multiplicities retained in the reduction. For H, the invertible system
\eqref{eq:HDorbit} determines a coupling locus in each dimension.
For P, the remainders \eqref{eq:Pdimensionremainders} exclude every other physical integer dimension for the profile \eqref{eq:PDfields}.
The finite integer-power classification extends this comparison to the additional profiles specified in Appendix~\ref{app:Pcoefficients}.

\section{Horizon functionals and finite-section charges}
\label{sec:charges}

Time translations act on the future-regular scalar through $\mathcal L_\xi\phi=q$, and the combined transformation
$(\xi,\delta_{\rm shift}\phi=-q)$ preserves the complete configuration.
We derive its surface-charge balance from the action \eqref{eq:action}, including the cubic scalar interaction, and evaluate
the curvature functional, Noether charge and Hamiltonian variation on the two fixed-action branches \cite{Wald1993,IyerWald1994}. The planar variation follows the source--temperature ray
\eqref{eq:PsourceT}.

\subsection{Covariant potentials}

Write $\mathcal N=(16\pi G_5)^{-1}$ and let $\mathcal L$ denote the scalar Lagrangian in parentheses in \eqref{eq:action}, excluding $\mathcal N$. The Lagrangian five-form is $\mathbf L=\mathcal N\mathcal L\boldsymbol\epsilon$. For the oriented
volume form $\boldsymbol\epsilon$, set $\boldsymbol\epsilon_a=i_{\partial_a}\boldsymbol\epsilon$ and $\boldsymbol\epsilon_{ab}=i_{\partial_b}i_{\partial_a}\boldsymbol\epsilon$,
with $\epsilon_{vr123}>0$. The metric variation is
$\eta_{ab}=\delta g_{ab}$, with
$\eta^{ab}=g^{ac}g^{bd}\eta_{cd}=-\delta g^{ab}$ and $\eta=g^{ab}\eta_{ab}$. Antisymmetrization has weight one half.

At fixed metric and scalar covariant derivatives, the Riemann derivative is
\begin{equation}
\begin{aligned}
\mathcal P^{abcd}
&=\frac{\partial\mathcal L}{\partial R_{abcd}}
=g^{a[c}g^{d]b}+\alpha_1\Pi^{abcd}+\alpha_2\mathcal D^{abcd},\\
\Pi^{abcd}
&=2R^{abcd}
-4\left(g^{a[c}R^{d]b}-g^{b[c}R^{d]a}\right)
+2R\,g^{a[c}g^{d]b},\\
\mathcal D^{abcd}
&=\frac14\left(g^{ac}\nabla^b\phi\nabla^d\phi-g^{ad}\nabla^b\phi\nabla^c\phi
-g^{bc}\nabla^a\phi\nabla^d\phi+g^{bd}\nabla^a\phi\nabla^c\phi\right)
-\frac X2g^{a[c}g^{d]b}.
\end{aligned}
\label{eq:PHfullP}
\end{equation}
With the shift current \eqref{eq:current}, direct variation of the
action gives
\begin{equation}
\begin{aligned}
\boldsymbol\Theta&=\mathcal N\Theta^a\boldsymbol\epsilon_a,\\
\Theta^a
&=2\mathcal P^{abcd}\nabla_d\eta_{bc}
-2(\nabla_d\mathcal P^{abcd})\eta_{bc}+J^a\delta\phi\\
&\quad+\alpha_3X\left(\nabla^a\delta\phi
-(\nabla_b\phi)\eta^{ab}+\frac12(\nabla^a\phi)\eta\right).
\end{aligned}
\label{eq:PHfullTheta}
\end{equation}
The last line follows from varying the Hessian in $X\Box\phi$ and integrating the connection variation once. It is the explicit cubic contribution to the boundary potential.

For a field-independent vector $\xi^a$, the corresponding charge is
\begin{equation}
\begin{aligned}
\mathbf Q_\xi&=\frac{\mathcal N}{2}Q_\xi^{ab}
\boldsymbol\epsilon_{ab},\\
Q_\xi^{ab}
&=-2\mathcal P^{abcd}\nabla_c\xi_d
+4\xi_d\nabla_c\mathcal P^{abcd}
+\alpha_3X\bigl((\nabla^a\phi)\xi^b-(\nabla^b\phi)\xi^a\bigr).
\end{aligned}
\label{eq:PHfullQ}
\end{equation}
The kinetic and quartic scalar terms enter through $J^a$, while the cubic term contributes explicitly to both potentials. On shell,
\begin{equation}
\mathbf J_\xi
=\boldsymbol\Theta(\mathcal L_\xi g,\mathcal L_\xi\phi)
-i_\xi(\mathcal N\mathcal L\boldsymbol\epsilon)
=\dd\mathbf Q_\xi .
\label{eq:PHNoetheridentity}
\end{equation}
Define the shift-current form
$\mathbf j=\mathcal N J^a\boldsymbol\epsilon_a$. For $\xi=\partial_v$ on either background, this identity becomes
$\mathbf J_\xi=q\mathbf j-i_\xi(\mathcal N\mathcal L\boldsymbol\epsilon)$.

\subsection{Future-horizon data and the curvature functional}

Let $\mathcal C$ be a spatial section of the future Killing horizon, with area $A_{\mathcal C}$, induced metric $\sigma_{ij}$, intrinsic curvature $\mathcal R[\sigma]$, and binormal $e_{ab}e^{ab}=-2$.
The Einstein and Einstein-tensor terms in \eqref{eq:PHfullP} have
the binormal contractions
\begin{equation}
g^{a[c}g^{d]b}e_{ab}e_{cd}=-2,\qquad
\mathcal D^{abcd}e_{ab}e_{cd}
=\sigma^{ab}\nabla_a\phi\nabla_b\phi .
\label{eq:PHbinormal}
\end{equation}
The second contraction cancels the normal scalar gradients between the two curvature terms in $G^{ab}\phi_a\phi_b$.
On these nonexpanding horizon sections, the extrinsic terms in the contracted Gauss relation vanish, so
$\Pi^{abcd}e_{ab}e_{cd}=-4\mathcal R[\sigma]$
\cite{JacobsonMyers1993}. These contractions give
\begin{equation}
\begin{aligned}
S_{\rm curv}
&=-\frac1{8G_5}\int_{\mathcal C}\dd^3x\sqrt{\sigma}\,
\mathcal P^{abcd}e_{ab}e_{cd}\\
&=\frac1{4G_5}\int_{\mathcal C}\dd^3x\sqrt{\sigma}\,
\left[1+2\alpha_1\mathcal R[\sigma]
-\frac{\alpha_2}{2}D_i\phi D^i\phi\right].
\end{aligned}
\label{eq:PHcurvaturefunctional}
\end{equation}
Since $D_i\phi=0$ on both branches,
\begin{equation}
S_{{\rm curv},P}=\frac{A_P}{4G_5},\qquad
S_{{\rm curv},H}=\frac{A_H}{4G_5}
\left(1-\frac{12\alpha_1}{R_H^2}\right).
\label{eq:PHcurvaturevalues}
\end{equation}
For these stationary-up-to-shift configurations, we evaluate the Hamiltonian variation directly on regular future-horizon sections through the covariant potentials, keeping the scalar contribution explicit \cite{FengLiuLuPope2015,MinamitsujiMaeda2023}.

\subsection{The shift current in the surface-charge balance}

For the fixed contravariant vector $\xi=\partial_v$, define the surface-charge variation by $\mathbf k_\xi=\delta\mathbf Q_\xi-i_\xi\boldsymbol\Theta$.
A linearized solution $\delta$ at fixed couplings obeys the covariant symplectic identity
\begin{equation}
\dd\mathbf k_\xi
=\boldsymbol\omega(\delta,\mathcal L_\xi),\qquad
\boldsymbol\omega(\delta,\mathcal L_\xi)
=\delta\boldsymbol\Theta(\mathcal L_\xi)
-\mathcal L_\xi\boldsymbol\Theta(\delta).
\label{eq:PHsymplecticidentity}
\end{equation}
On the stationary-up-to-shift family,
$\boldsymbol\Theta(\mathcal L_\xi)=q\mathbf j$.
All coefficients in the potential are stationary; its only explicit $v$ dependence comes from
$J^a\delta\phi=J^a(v\,\delta q+\delta p)$.
Consequently,
\begin{equation}
\mathcal L_\xi\boldsymbol\Theta(\delta)=\delta q\,\mathbf j,
\qquad
\boldsymbol\omega
=\delta(q\mathbf j)-\delta q\,\mathbf j=q\,\delta\mathbf j .
\label{eq:PHshiftbalanceproof}
\end{equation}
The $\delta q$ terms cancel, giving the balance
\begin{equation}
\mathbf k_\xi=\delta\mathbf Q_\xi-i_\xi\boldsymbol\Theta,\qquad
\dd\mathbf k_\xi=q\,\delta\mathbf j .
\label{eq:PHchargeidentity}
\end{equation}
Thus the varying bulk shift-charge density determines the change of the surface-charge variation, including along families with varying $q$. The same argument applies in every dimension to a shift-invariant covariant potential of this form and a stationary-up-to-shift family.

Integrating over a region $\Sigma$ with fixed boundary gives
\begin{equation}
\int_{\partial\Sigma}\mathbf k_\xi=q\,\delta C_\Sigma,
\qquad C_\Sigma=\int_\Sigma\mathbf j.
\label{eq:PHglobalshiftbalance}
\end{equation}
The global scalar shift carries the volume charge $C_\Sigma$.
The combined time translation and shift $c=-q$ preserves the background; for varying $q$, its field-dependent generator gives the adjusted Hamiltonian variation $\delta H_\xi-q\,\delta C_\Sigma$. Here $\delta H_\xi$ denotes the boundary variation in
\eqref{eq:PHglobalshiftbalance}.

For a homogeneous section at fixed $r,v$, define $K_\xi(r)$ by $\left.\mathbf k_\xi\right|_{r,v}=K_\xi(r)\dd V_{\gamma_k}$, where $\dd V_{\gamma_k}$ is the unit-fiber volume form. Our orientation gives
\begin{equation}
K_\xi(r)=\mathcal N\left[\delta(r^3Q_\xi^{vr})
+r^3\Theta^r\right].
\label{eq:PHcutcharge}
\end{equation}
We take the variation at fixed EF coordinates and then pull back the result to the background horizon.

\subsection{Radial potentials and the continuous planar family}

Set $n=1$, $\Sigma=r$, and write
\begin{equation}
w=p',\qquad s=q+fw,\qquad X=2qw+fw^2,\qquad
\Box\phi=s'+\frac{3s}{r}.
\label{eq:PHchargeblocks}
\end{equation}
In these variables, the scalar Lagrangian reduces to
\begin{equation}
\begin{aligned}
r^3\mathcal L&=r^3\left[
R+\alpha_1\GB+(\alpha_0+\alpha_2A)X
+\alpha_3X\Box\phi+\alpha_4X^2\right],\\
A&=\frac{3f'}{2r}+\frac{3(f-k)}{r^2},\\
R&=-f''-\frac{6f'}r+\frac{6(k-f)}{r^2},\\
\GB&=\frac{12(f')^2}{r^2}
-\frac{24(k-f)f'}{r^3}
-\frac{12(k-f)f''}{r^2}.
\end{aligned}
\label{eq:PHchargeLred}
\end{equation}
At fixed EF coordinates, $\eta_{vv}=-\delta f$ is the only nonzero
metric variation, while
\begin{equation}
\delta\phi=v\,\delta q+\delta p,\qquad
\nabla^r\delta\phi=\delta q+f\,\delta p'.
\label{eq:PHchargevariation}
\end{equation}
Substituting these variations into \eqref{eq:PHfullTheta} and
\eqref{eq:PHfullQ} gives
\begin{equation}
\begin{aligned}
r^3\Theta^r={}&-\left[r^3+12\alpha_1r(k-f)\right]\delta f'\\
&+\left[-3r^2+12\alpha_1\{rf'-(k-f)\}
+\frac32\alpha_2r^2X+\alpha_3r^3Xw\right]\delta f\\
&+r^3J^r(v\,\delta q+\delta p)
+\alpha_3r^3X(\delta q+f\,\delta p'),\\
Q_\xi^{vr}={}&f'\left[1+\frac{12\alpha_1(k-f)}{r^2}\right]
-\frac{3\alpha_2}{r}s^2-\alpha_3Xs .
\end{aligned}
\label{eq:PHradialpotentials}
\end{equation}
These expressions retain the variation of the boundary time slope along with the radial scalar and metric variations.

For P, set $a=r_h$ and $\gamma=q/a$, with $L,\gamma$ fixed by the action. The orbit \eqref{eq:Porbit} then fixes
\begin{equation}
\begin{gathered}
s=\frac{\gamma a^2}{r},\qquad
X=-\frac{\gamma^2a^2L^2}{r^2},\qquad
\delta f=-\frac{2a}{L^2}\delta a,\qquad \delta f'=0,\\
\delta q=\gamma\delta a,\qquad
\delta p'=-\frac{\gamma L^2}{(r+a)^2}\delta a .
\end{gathered}
\label{eq:Pfixedactionvariation}
\end{equation}
Substitution into \eqref{eq:PHradialpotentials} yields
\begin{equation}
Q_{\xi,P}^{vr}=-\frac{10r}{L^2}+\frac{12a^2}{L^2r},\qquad
r^3\Theta_P^r=\frac{-30ar^2+6a^3}{L^2}\,\delta a .
\label{eq:Pchargepotential}
\end{equation}
The Einstein-tensor and cubic contributions to the background $Q_\xi^{vr}$ cancel on this branch; both interactions contribute to the full variation.

The on-shell Lagrangian satisfies the Noether identity:
\begin{equation}
\begin{aligned}
\mathcal L_P&=\frac{40}{L^2}-\frac{36a^2}{L^2r^2},\\
\frac1{r^3}\partial_r(r^3Q_{\xi,P}^{vr})
&=qJ_P^v-\mathcal L_P
=-\frac{40}{L^2}+\frac{24a^2}{L^2r^2}.
\end{aligned}
\label{eq:Pchargeclosure}
\end{equation}
At fixed scalar zero mode, the temporal boundary potential is
\begin{equation}
r^3\Theta_P^v=6a\left(1-\frac{3r}{r+a}
-\frac{2rv}{L^2}\right)\delta a.
\label{eq:Ptemporalpotential}
\end{equation}
A variation of the constant zero mode adds $r^3J_P^v\delta\phi_0$ to this expression and leaves the radial section charge unchanged.
The radial and temporal boundary derivatives are
\begin{equation}
\begin{aligned}
\delta(r^3\mathcal L_P)&=-\frac{72ar}{L^2}\delta a,\\
\partial_r(r^3\Theta_P^r)&=-\frac{60ar}{L^2}\delta a,\qquad
\partial_v(r^3\Theta_P^v)
=r^3J_P^v\delta q=-\frac{12ar}{L^2}\delta a .
\end{aligned}
\label{eq:Pboundaryclosure}
\end{equation}
Their sum reproduces the complete on-shell variation, including the contribution from the varying time slope.

Combining the radial potential and boundary term in
\eqref{eq:PHcutcharge} gives
\begin{equation}
K_{\xi,P}(r)=\frac{6\mathcal N a(a^2-r^2)}{L^2}\,\delta a,\qquad
K_{\xi,P}(a)=0 .
\label{eq:Pchargevariation}
\end{equation}
For any two fixed exterior sections,
\begin{equation}
K_{\xi,P}(r_2)-K_{\xi,P}(r_1)
=-\frac{6\mathcal N a}{L^2}(r_2^2-r_1^2)\delta a
=q\int_{r_1}^{r_2}\delta\!\left(\mathcal N r^3J_P^v\right)\dd r .
\label{eq:Pfinitebalance}
\end{equation}
Here $r^3J_P^v=-12ar/(L^2\gamma)$ supplies the bulk term directly.
The bare volume charge grows quadratically with the outer radius, while \eqref{eq:Pfinitebalance} gives its exact finite-section balance. The curvature functional has the variation
\begin{equation}
T_P\,\delta\!\left(\frac{S_{{\rm curv},P}}{V_3}\right)
=\frac{6\mathcal N a^3}{L^2}\delta a\ne K_{\xi,P}(a),
\label{eq:Pentropycontrast}
\end{equation}
where $V_3$ is the regulated planar volume. Equations~\eqref{eq:Pfinitebalance}
and \eqref{eq:Pentropycontrast} determine the distinct contributions of the source variation and the horizon geometry along the thermal ray.

\subsection{The isolated hyperbolic solutions}

For H, define the dimensionless constant
\begin{equation}
C_H=2-\frac{24\alpha_1}{R_H^2}
-3\alpha_2q^2-\alpha_3q^3R_H .
\label{eq:Hchargeconstant}
\end{equation}
The full background expressions are
\begin{equation}
Q_{\xi,H}^{vr}=\frac r{R_H^2}C_H,\qquad
J_H^a=0,\qquad \mathcal L_H=-\frac{4C_H}{R_H^2}.
\label{eq:Hbackgroundcharge}
\end{equation}
They satisfy $\nabla_bQ_{\xi,H}^{ab}=-\xi^a\mathcal L_H$.
At either nondegenerate root \eqref{eq:Hcommonroots}, varying $M=\mathcal S=0$ at fixed couplings and fixed horizon quotient gives $\delta R_H=\delta\chi=\delta q=0$ through $\Delta_H\ne0$. The remaining constant scalar shift has $\Theta^a=J_H^a\delta\phi_0=0$ and $\delta\mathbf j=0$, and hence $\mathbf k_\xi=0$.

The two branches thus realize the surface-charge identity through a varying bulk shift charge and an isolated scalar dressing, respectively.
For a covariant four-form $\boldsymbol\mu$ and fixed $\xi$, the identity $\delta(i_\xi\boldsymbol\mu)=i_\xi\delta\boldsymbol\mu$ cancels the boundary-term shifts of $\delta\mathbf Q_\xi$ and
$i_\xi\boldsymbol\Theta$. The Hamiltonian variation $\mathbf k_\xi$ is therefore invariant under $\mathbf L\mapsto\mathbf L+\dd\boldsymbol\mu$
\cite{JacobsonKangMyers1994}.

\section{Conclusions}
\label{sec:conclusions}

We have constructed a continuous planar family and two isolated hyperbolic black holes in one fixed Lovelock--Horndeski theory. The planar geometry is Weyl curved and has a free horizon scale; the hyperbolic solutions share a locally AdS metric with distinct scalar dressings. Their dimensional continuations are fixed by the nonlinear field equations: the hyperbolic construction extends to $D\geq5$, whereas the displayed planar profile exists only in $D=5,7$. The scalar also changes the covariant charge balance. Time translations act through a global shift, yielding $\dd\mathbf k_\xi=q\,\delta\mathbf j$. Along the planar family, the varying bulk shift-charge density supplies this balance despite zero radial flux and zero horizon charge variation. At the hyperbolic roots, the fixed-action constraints leave no horizon-scale tangent.
These results determine both the allowed scalar dressings and their charge variations in a theory motivated by heterotic compactification \cite{WuStone2026}. With the boundary variational problem supplied by holographic renormalization \cite{WuHolographicRenormalization2026}, the exact backgrounds provide data for renormalized thermal responses \cite{deHaroSkenderisSolodukhin2001} and for the coupled characteristic analysis \cite{ReallTanahashiWay2014}.

\appendix
\section{Boundary terms and the horizon Noether charge}
\label{app:charges}

On a simple horizon at $r=a$, $s(a)=q$ and $\kappa=f'(a)/2$.
Define the rescaled horizon Noether charge
$S_Q=(2\pi/\kappa)\int_{\mathcal C}\mathbf Q_\xi$.
Its difference from the curvature functional is
\begin{equation}
S_Q-S_{\rm curv}
=-\frac{A_{\mathcal C}}{4G_5f'(a)}
\left[\frac{3\alpha_2q^2}{a}+\alpha_3X(a)q\right].
\label{eq:PHrawchargedifference}
\end{equation}
The scalar contributions cancel on P. On H,
\begin{equation}
S_{Q,H}=\frac{A_H}{4G_5}
\left[1-\frac{12\alpha_1}{R_H^2}
-\frac32\alpha_2q^2-\frac12\alpha_3q^3R_H\right].
\label{eq:Hrawcharge}
\end{equation}
At the common-action roots $\chi_\pm$ in \eqref{eq:Hcommonroots},
the dimensionless multipliers of both functionals are
\begin{equation}
\frac{4G_5S_{Q,H,\pm}}{A_H}
=1-3\sqrt2\mp3\sqrt{\frac12+\frac1{\sqrt2}},\qquad
\frac{4G_5S_{{\rm curv},H}}{A_H}=-2-3\sqrt2 .
\label{eq:Hrawrootvalues}
\end{equation}
Under a change of representative $\mathbf L\mapsto\mathbf L+\dd\boldsymbol\mu$,
with
\begin{equation}
\boldsymbol\mu=\mathcal N\beta\,{*}\dd\phi
=\mathcal N\beta\,\nabla^a\phi\,\boldsymbol\epsilon_a,\qquad
\beta=\text{constant},
\label{eq:PHboundaryrepresentative}
\end{equation}
the bulk equations are unchanged, while the potentials transform as
\begin{equation}
\boldsymbol\Theta\longmapsto\boldsymbol\Theta+\delta\boldsymbol\mu,
\qquad
\mathbf Q_\xi\longmapsto\mathbf Q_\xi+i_\xi\boldsymbol\mu,
\qquad
\mathbf k_\xi\longmapsto\mathbf k_\xi
\label{eq:PHrepresentativeinvariance}
\end{equation}
for fixed $\xi$. On a horizon section,
\begin{equation}
\int_{\mathcal C}i_\xi\boldsymbol\mu
=-\mathcal N\beta A_{\mathcal C}q,\qquad
\Delta S_Q=-\frac{\beta A_{\mathcal C}q}{8G_5\kappa}.
\label{eq:PHrawambiguity}
\end{equation}
Thus the boundary term shifts $S_Q$ while preserving the Hamiltonian
variation $\mathbf k_\xi$. The displayed P and H charges use the
Lagrangian representative \eqref{eq:action}.

\section{Dimensional compatibility and integer-power solutions}
\label{app:Pcoefficients}

\subsection{The thirteen dimensional coefficients}

The coefficient problem can be written entirely in the variables
$b_i$ of \eqref{eq:Pdimensionless}. Set $x=r/r_h$ and normalize the
metric and scalar scales in the radial equations. This amounts to using
$L=r_h=q=1$ in the coefficient calculation, with $\alpha_i$ replaced by
$b_i$; restoring the scales gives \eqref{eq:Pdimensionless}.
Define
\begin{equation}
 A_\nu=(\nu-1)(\nu-2)b_1,\qquad
 c_{Q,j}=[x^j]\mathcal P_Q,
 \label{eq:Pcoefficientdefinition}
\end{equation}
where the normalized $\mathcal P_Q$ are obtained from
\eqref{eq:Pscaledresiduals}. The lapse polynomial has degree five,
but $c_{n,5}=c_{n,4}$. Retaining one of these identical rows gives
the following thirteen coefficients:
\begin{align}
c_{n,0}&=-2\nu(\nu-3)A_\nu+\nu(\nu-5)b_2+4b_3-2b_4,\notag\\
c_{n,1}&=-2\nu(\nu-3)A_\nu+\nu(3\nu-7)b_2
-4(\nu-1)b_3+6b_4,\notag\\
c_{n,2}&=4\nu(\nu-1)A_\nu+\nu(\nu+1)b_2+2b_0-4b_3
-2\nu(\nu-1),\notag\\
c_{n,3}&=4\nu(\nu-1)A_\nu-\nu(\nu+1)b_2-2b_0
-2\nu(\nu-1),\notag\\
c_{n,4}&=-2\nu(\nu+1)(A_\nu-1),
\label{eq:PcoeffN}\\
c_{f,0}&=(\nu+1)(\nu b_2-2b_3)+4b_4,\notag\\
c_{f,1}&=4(\nu b_2-b_3),\qquad
c_{f,2}=-2b_0-\nu(\nu-1)b_2,
\label{eq:PcoeffF}\\
c_{\Sigma,0}&=2(\nu-3)(\nu-4)A_\nu
-(\nu-1)(\nu-6)b_2-4b_3+2b_4,\notag\\
c_{\Sigma,2}&=-4(\nu-1)(\nu-2)A_\nu
-\bigl[(\nu-1)(\nu-2)+2\bigr]b_2-2b_0
+2(\nu-1)(\nu-2),\notag\\
c_{\Sigma,4}&=2\nu(\nu+1)(A_\nu-1),
\label{eq:PcoeffSigma}\\
c_{p,0}&=(\nu-3)\bigl[\nu(\nu-1)b_2-2\nu b_3+4b_4\bigr],\notag\\
c_{p,2}&=-(\nu-1)\bigl[\nu(\nu+1)b_2+2b_0-2b_3\bigr].
\label{eq:PcoeffPhi}
\end{align}
These rows retain the four Euler equations and their algebraic
relations. In particular,
$c_{\Sigma,4}=-c_{n,4}$. At $\nu=3$, $c_{p,0}$ vanishes identically.
The remaining twelve rows reduce, after constant rescaling and
removal of repetitions, to the nine polynomials in \eqref{eq:Prows}.

For $\nu=2$, the lapse row gives $c_{n,4}=12$, excluding that dimension
without dividing by $\nu-2$. 
For $\nu>2$, the coefficient matrix of $(c_{n,4},c_{f,1},c_{f,2},c_{f,0},c_{n,2})$, with columns ordered
as $(b_0,b_1,b_2,b_3,b_4)$, has the nonzero determinant \eqref{eq:Ppivotminor}. The first four equations give
\begin{equation}
A_\nu=1,\qquad b_3=\nu b_2,\qquad
b_0=-\frac{\nu(\nu-1)}2b_2,\qquad
b_4=\frac{\nu(\nu+1)}4b_2.
\label{eq:Pfirstfourpivots}
\end{equation}
The fifth then becomes
\begin{equation}
c_{n,2}=2\nu\bigl[(\nu-1)-b_2\bigr],
\label{eq:Plastpivot}
\end{equation}
and hence fixes the entire orbit \eqref{eq:Pdimensionorbit}.
Substitution into all thirteen rows leaves only
\begin{equation}
c_{n,0}=c_{n,1}=\frac{\nu(\nu-3)(\nu-5)}2,\qquad
c_{\Sigma,0}=-\frac{(\nu-3)(\nu-4)(\nu-5)}2.
\label{eq:Pexplicitremainders}
\end{equation}
This establishes necessity at every physical integer dimension and
sufficiency at $\nu=3,5$. Since the factors used in
\eqref{eq:Pscaledresiduals} are nonzero for $r\geq r_h>0$, vanishing
of the coefficient polynomials is equivalent to vanishing of the
uncollected radial equations throughout the exterior.

\subsection{Algebraic exclusion of other integer-power profiles}

We classify the integer-power profiles \eqref{eq:Ptrial} with
$1\leq m,z\leq5$ in horizon dimensions $\nu=2,\ldots,7$.
The endpoint condition fixes $c=-q$. In the dimensionless variables
used above, the radial functions are
\begin{equation}
 F=x^2-x^{2-m},\qquad
 w\equiv p'=-\frac{1-x^{-z}}{F},\qquad
 w'=\frac{(1-x^{-z})F'-zx^{-z-1}F}{F^2}.
 \label{eq:Pscanfunctions}
\end{equation}
The four Euler expressions involve radial derivatives up to $F''$
and $w'$. Their normalized values are affine in the five couplings:
\begin{equation}
\varepsilon_Q(x)\equiv x^{-\nu}\E_Q(x)
=\sum_{i=0}^4 A_{Q i}(x;\nu,m,z)b_i+d_Q(x;\nu,m,z).
\label{eq:Pscanaffine}
\end{equation}

A five-dimensional example exhibits the incompatibility directly.
For $(\nu,m,z)=(3,1,1)$, the functions reduce to
$F=x^2-x$ and $w=-x^{-2}$. Variation of the reduced action before
imposing these profiles gives the lapse equation
\begin{equation}
\begin{aligned}
\varepsilon_n(x)={}&24b_1-12+\frac{9-36b_1}{x}
+\frac{b_0+12b_1+6b_2}{x^2}\\
&+\frac{-b_0-\tfrac{21}{2}b_2+2b_3}{x^3}
+\frac{\tfrac92b_2+5b_3-3b_4}{x^4}
-\frac{3b_3+2b_4}{x^5}+\frac{b_4}{x^6}.
\end{aligned}
\label{eq:Pscananalyticexample}
\end{equation}
The constant and $x^{-1}$ coefficients require $b_1=1/2$ and
$b_1=1/4$, respectively. Equivalently, they satisfy the
coupling-independent relation
\begin{equation}
\frac32[x^0]\varepsilon_n+[x^{-1}]\varepsilon_n=-9.
\label{eq:Pscananalyticcontradiction}
\end{equation}
The AdS term and the linear correction therefore impose incompatible
conditions on the Gauss--Bonnet coupling. The scalar couplings enter
at lower powers and cannot remove this incompatibility.

For the full finite classification, exact evaluation of
\eqref{eq:Pscanaffine} at exterior radii gives a necessary linear
system. Its entries are rational for integer $(\nu,m,z)$ and rational
$x>1$. We use the six equations in the order
\begin{equation}
(n,2),\ (f,2),\ (\Sigma,2),\ (p,2),\ (n,3),\ (f,3),
\label{eq:Pscanpoints}
\end{equation}
and denote their coefficient matrix and constant vector by
$A$ and $\boldsymbol d$. For each excluded profile, exact rational
elimination gives
\begin{equation}
\boldsymbol\omega^{\mathsf T}A=0,\qquad
\boldsymbol\omega^{\mathsf T}\boldsymbol d=1,
\qquad
\sum_j\omega_j\varepsilon_{Q_j}(x_j)=1.
\label{eq:Pscancontradiction}
\end{equation}
An exact radial solution would set every term in the sum to zero.
This identity therefore excludes the profile throughout the coupling
space. Five equations suffice at $\nu=2$; the other excluded cases
use all six. The complete rational matrices and elimination
identities are provided in the accompanying algebraic calculation.

Exact elimination excludes every exponent pair at $\nu=2,4,6,7$.
At $\nu=3$ and $\nu=5$, the unique admissible pair is $(m,z)=(2,1)$,
with couplings fixed by \eqref{eq:Pdimensionorbit}. The residual
polynomial coefficients \eqref{eq:Pexplicitremainders} vanish in both
dimensions, so the full radial equations hold at every exterior radius.
Together with the exclusion identities, this completes
the classification for $1\leq m,z\leq5$ and $2\leq\nu\leq7$.

\begingroup
\raggedright

\endgroup

\end{document}